\documentclass[aps,prd,amsmath,amssymb,nofootinbib,showpacs,reprint]{revtex4-1}

\usepackage{amsmath}
\usepackage{amsfonts}
\usepackage{amssymb}
\usepackage{bbm}	
\usepackage{slashed}

\usepackage{mathrsfs}
\usepackage[english]{babel}
\usepackage{microtype}
\usepackage[pdftex]{hyperref}
\usepackage{nicefrac}
\usepackage{graphicx}
\usepackage[babel]{csquotes}
\DeclareMathAlphabet{\boldmathe}{T1}{cmr}{bx}{it}
\newcommand{\mbf}[1]{\boldmathe{#1}}

\newcommand{\N}{{\mathbb{N}}}
\newcommand{\Z}{{\mathbb{Z}}}

\newcommand{\tr}{\operatorname{tr}}
\newcommand{\ii}{\mathrm{i}}
\newcommand{\D}{\mathrm{d}}
\newcommand{\cO}{\mathcal{O}}

\newcommand{\vev}[1]{\left\langle #1 \right\rangle}

\newcommand{\Nr}{N}
\newcommand{\Nrc}{N^\text{c}}
\newcommand{\Ni}{N_\text{ir}}
\newcommand{\trd}{\tr_\mathrm{\hskip0.3mm D}}
\newcommand{\trf}{\tr_\mathrm{\hskip0.3mm F}}
\newcommand{\lamth}{\lambda}
\newcommand{\Sigmam}{\Sigma_\mathrm{max}}
\def\vx{\mbf{x}}
\def\vy{\mbf{y}}
\usepackage{color}

\newcommand{\iFT}{\operatorname{FT}^{-1}}

\newcommand{\inputFig}[1]{\includegraphics{#1.pdf}}
\graphicspath{{./}}

\begin{document}
\title{Absence of chiral symmetry breaking in Thirring models in 1+2 dimensions}

\newcommand{\FSU}{Theoretisch-Physikalisches Institut, Friedrich-Schiller-Universit{\"a}t Jena, 
07743 Jena, Germany}

\author{Julian J. Lenz}
\email{julian.johannes.lenz@uni-jena.de}
\affiliation{\FSU}

\author{Bj\"orn H. Wellegehausen}
\email{bjoern.wellegehausen@uni-jena.de}
\affiliation{\FSU}

\author{Andreas Wipf\,}
\email{wipf@tpi.uni-jena.de}
\affiliation{\FSU}

\begin{abstract}
The Thirring model is an interacting fermion theory with
current-current interaction.
 The model in $1+2$ dimensions
has applications in condensed-matter physics to describe the
electronic excitations of Dirac materials.
Earlier investigations with Schwinger-Dyson equations, 
the functional
renormalization group and lattice simulations with staggered fermions suggest that a critical number of (reducible) 
flavors $\Nrc$ exists, below which chiral symmetry can be 
broken spontaneously. Values for $\Nrc$ found in the literature 
vary between $2$ and $7$.
Recent lattice studies with chirally invariant SLAC
fermions have indicated that chiral symmetry is unbroken for 
all integer flavor numbers \cite{Wellegehausen:2017goy,phd_schmidt}. 
An independent simulation based on domain wall fermions 
seems to favor a critical flavor-number
that satisfies $1<\Nrc<2$ \cite{Hands:2018vrd}. However,
in the latter simulations difficulties in reaching
the massless limit in the broken phase 
(at strong coupling and after the $L_s\to\infty$ limit
has been taken) are encountered. To find an accurate
value $\Nrc$ we study the Thirring model
(by using an analytic continuation of the parity
even theory to arbitrary real $\Nr$) for $\Nr$ 
between $0.5$ and $1.1$. We investigate the
chiral condensate, the spectral
density of the Dirac operator, the spectrum
of (would-be) Goldstone bosons and 
the variation of the filling-factor and
conclude that the critical
flavor number is $\Nrc=0.80(4)$. 
Thus we see no
chiral symmetry breaking in all Thirring models with
$1$ or more flavors of ($4$-component) fermions. 
Besides the transition to the 
unphysical lattice artifact phase we find 
strong evidence for a hitherto unknown phase
transition that exists for $\Nr>\Nrc$
and should answer the question of where to
construct a continuum limit.
\end{abstract}
\maketitle
\section{Introduction}
\noindent
The Thirring model \cite{Thirring1958}
in $2$ space-time dimensions 
is integrable and in the massless
limit even soluble \cite{Klaiber:1967jz,Sachs:1995dm}.
The model in $3$ space-time dimensions is of interest for various
reasons, e.g.\@ its close relationship to
QED$_3$
\cite{Hands1995,Itoh1995,Gusynin2016,Kotikov2016a} or its
relevance in solid state physics, where it
describes low-energy electronic properties of materials 
like graphene \cite{Semenoff1984,Hands2008} or 
high-temperature superconductors~\cite{Herbut2002,Franz2002}. 
In $3$ dimensions the model is perturbatively non-renormalizable
but can be renormalized in the limit of large flavor numbers $\Nr$
\cite{Parisi1975,Gawedzki:1985ed,Rosenstein:1990nm,Hands1995}. 
Thus it provides a simple realization \cite{Braun:2010tt} of 
the concept of asymptotic safety \cite{Weinberg:1976xy}.
In the large-$N$
limit one finds an unbroken U$(2\Nr)$ symmetry for every 
coupling strength. On the other hand, in the limit 
$\Nr=\nicefrac{1}{2}$ the Thirring model is equivalent to the
Gross-Neveu model. The latter exhibits (for all
$N$) a second order phase transition from a symmetric 
gapless (massless) phase 
at weak coupling to a spontaneously broken gapped (massive) 
phase at sufficiently  strong couplings\footnote{More precisely, 
the Thirring model
with $1$ irreducible $2$\emph{-component} Fermion flavor 
is the same as the Gross-Neveu model with $1$ irreducible
Fermion flavor.}. We conclude that the Thirring model
exhibits no chiral phase transition for large $N$ but shows
a second order phase transition at $\Nr=\nicefrac{1}{2}$.
The question about the critical flavor number $\Nrc$ below
which the Thirring model shows a chiral phase transition 
has been intensively discussed in the past.
While early results obtained with functional methods or
staggered lattice fermions range from $\Nrc=2$ to $\Nrc=\infty$
\cite{Gomes1991,Itoh1995, Sugiura1997,Kondo1995,Gies2010,
Janssen2012a,Kim1996,DelDebbio1996,DelDebbio1997,DelDebbio1999,
Hands1999,Christofi2007}, more recent lattice studies 
with chiral fermions favor smaller values of $\Nrc$. 
In particular,  based on simulations with 
massless (chiral) fermions we argued 
that the U$(2\Nr)$-\,symmetry is unbroken for all integer flavor numbers $\Nr\gtrapprox 1$ \cite{Wellegehausen:2017goy}. 
For $\Nr=1$ the effective potential for 
the chiral condensate is almost flat at
the origin such that we could not completely rule out the
possibility that there is SSB for $N=1$. 
In an interesting recent work Simon Hands applied 
domain-wall fermions to study the chiral condensate
and masses of the (would-be) Goldstone bosons
\cite{Hands:2018vrd}. 
The results support our finding that $\Nrc$ is
smaller than hitherto believed with the notable difference that 
he interprets his data as an evidence for $1<\Nrc<2$.
In a recent explorative
functional renormalization group (FRG) study with 
momentum-dependent couplings and the pseudo-spectral method
the critical behavior of four-fermion theories
\cite{Dabelow:2019sty} has been reconsidered. 
While a precise estimate for $\Nrc$ remains difficult in 
these elaborate FRG-studies, the new results are compatible 
with the lattice studies based on chiral fermions.

This work aims to solve the
discrepancy between the results obtained with domain-wall and
SLAC fermions. For that purpose we first
performed simulations
for $38$ different non-integer values of $\Nr$ between $0.5$ 
and $1.1$ and calculated the corresponding chiral condensates.
This way we already find strong evidence for 
a critical flavor number significantly lower 
than $1.0$. However, due to the computational 
cost of the algorithm a reliable extrapolation to infinite volume is difficult. But with the help 
of a careful
study of the (would-be) Goldstone spectrum and the spectral density of the Dirac operator we could not only assure unbroken symmetry at $\Nr\geq 1.0$ but also verify the proposed SSB
$\mathrm{U}(2) \to \mathrm{U}(1) \otimes \mathrm{U}(1)$. We conclude that indeed there is a critical flavor number $\Nrc\approx 0.80$ which 
is considerably smaller than $1.0$. A
similarly accurate value for $\Nrc$
comes from studying the susceptibility of the four
Fermi term in the Lagrangian which signals
-- besides the well-known transition to the artificial 
lattice phase -- a new interaction driven phase transition 
for all models with flavor
numbers $\Nr\geq N^\mathrm{t}=0.78(4)$.
We argue that $N^\mathrm{t}$ should be identified
with $\Nrc$. 
There is evidence that the new transition
is of second order and can be used to construct a
continuum limit of the lattice Thirring models.
Interestingly, this new transition seems 
unrelated to any change of symmetry.

To summarize: All results of our simulations 
with SLAC 
fermions consistently show that chiral symmetry is
not broken in all massless Thirring models with $N=1,2,3,\dots$ four-component fermions.

The paper is organized as follows: In the first section we recall 
relevant properties of the (reducible) Thirring model. 
For more details we refer to our earlier and much more
detailed work \cite{Wellegehausen:2017goy}, in which we
investigated Thirring models with irreducible $2$-\,component 
spinor-fields and with reducible $4$\,-component
fields. In the present work we focus on the reducible
and parity-even case considered in other works on 
the Thirring model in $3$ dimensions.
In the next two sections we present our lattice results 
for the chiral condensate and the spectral density -- 
from which we extract a first estimate of the critical 
flavor number. Then we discuss the correlation
matrix for interpolating operators 
for the scalar and pseudoscalar mesons.
The simulation results for the meson spectra support 
the proposed symmetry breaking pattern of 
chiral symmetry. In the following section 
we present our simulation results for the expectation value
of the interaction term $\propto (\bar{\psi}\Gamma_{\mu}\psi)^2$
and the corresponding susceptibility.
The expectation value is related to the mean
filling factor of the fermions.

In appendix A we prove some useful properties of the
spectral density and fermion Green function which
follow from parity invariance of the reducible theory.
Appendices B and C contain some technical details concerning 
numerical differentiation and our simulations.
	
\section{The Thirring model: order parameter and spectral density}
\label{chap:Method}
\noindent
The Lagrangian density of the Thirring model in 
three-dimensional Euclidean space-time has the form
	\begin{equation}
	\label{eq:lagrangian}
	\mathcal{L}=\sum_{a=1}^N\bar{\psi}_a\,\ii \Gamma^\mu\partial_\mu\,\psi_a
	-\frac{g^2}{2\Nr}j^\mu j_\mu,\;\;\;
	j^\mu=
	\sum_{a=1}^N\bar{\psi}_a\Gamma^{\mu}\psi_a\,,
	\end{equation}
and contains a vector-vector interaction built from
$\Nr$ flavors $\psi_1,\dots,\psi_N$.
In the present work $\psi_a$ (or $\psi$) always denotes
a $4$-component reducible spinor.
The hermitean matrices $\Gamma_\mu$ with $\mu=1,2,3$ 
form a $4$-dimensional \emph{reducible} representation of the
Clifford algebra. 
After introducing a Hubbard-Stratonovich auxiliary vector 
field $v^{\mu}$, a subsequent integration over the 
fermion fields leads to the partition function
(see \cite{Wellegehausen:2017goy} for more details)
	\begin{align}
		Z&=\int\mathcal{D}v^{\mu}\,
		e^{-S_\mathrm{eff}(v)},\quad\mathrm{with}\label{eff_action1}
		\\ 
		S_\mathrm{eff}
		&=\lamth\int \!\D^3x\, v^\mu v_\mu-\Nr\log\det (\ii D),\quad \lamth=\frac{N}{2g^2}\,.\nonumber
	\end{align}
Here we used that the determinant
for $N$ flavors is just the $N$'th power
of the determinant for $1$-flavor with
Dirac operator
\begin{equation}
	D=\Gamma^\mu \left(\partial_\mu-\ii v_\mu\right)+m\,.	\label{diracop1}
\end{equation} 
We introduced a chirality-breaking fermion mass
which is needed to control our lattice Monte-Carlo simulations
in the chirally broken phase. The eigenvalues of $\ii D$
come in pairs $(\lambda+\ii\,m,-\lambda+\ii\,m)$ such 
that the fermion determinant is real and positive,
 \begin{equation}
	\det(\ii D)
	=\left(\det D^\dagger D\right)^\frac{1}{2}\geq 0\,.
	\end{equation}	
This means that the effective action in (\ref{eff_action1})
is real or that the (massive or massless) Thirring model
with $N$ reducible flavors has no sign problem.
Hence in the well-known auxiliary field formulation 
the model can  be simulated by Monte-Carlo methods 
on a space-time lattice. At this point
we observe that $\Nr$ is just a parameter
that can be varied continuously. In the present work
we will focus on $\Nr\lessapprox 1$ and thus
consider lattice models which continuously
extrapolate to $\Nr=1.0$ from below. The so defined
models have no parity anomaly for any real $\Nr$.

The massless Thirring model with $N$ reducible flavors
is invariant  under the discrete $\mathbbm{Z}_2$ parity 
transformation as well as global U$(2\Nr)$ chiral transformations. 
These symmetries, together with the discrete $C$ and $T$ 
symmetries, are well explained in \cite{Janssen2012a}.
A technical problem here is that on a finite lattice
the condensates vanish in the massless case exactly for
every vector field configuration and a careful
extrapolation to vanishing fermion mass is difficult.

For performance reasons, we simulate the
theory in a $2$-component irreducible representation of the 
Clifford algebra. A convenient reducible representation is
\begin{equation}
\begin{aligned}
 \Gamma^\mu=&\sigma_3 \otimes \gamma^\mu\,,&\Gamma_4&=\sigma_2 \otimes \sigma_0\,,\\ \quad \Gamma_5=&\sigma_1 \otimes \sigma_0 \quad \text{and} &\Gamma_{45}&\equiv \ii\Gamma_4\Gamma_5=\sigma_3 \otimes \sigma_0\,,
\label{gammas}
\end{aligned}
 \end{equation}
and the corresponding Dirac operator (\ref{diracop1}) reads
\begin{equation}
D=
\begin{pmatrix}
\slashed{D}+m&0\\ 0&-\slashed{D}+m
\end{pmatrix},
\quad \slashed{D}=
\gamma^\mu(\partial_\mu-\ii v_\mu)\,.\label{diracop3}
\end{equation}
At this point we change the fermionic
variables, 
\begin{equation}
\psi_a\to \Gamma_{45}\psi_a,\quad 
\bar\psi_a\to\bar{\psi}_a,\quad a=1,\dots,N,
\label{changevar1}
\end{equation}
such that the Dirac operator $\slashed{D}$
(acting on two-component irreducible spinors) enters $D$
with the same sign\footnote{If $\slashed{D}$ would have 
a different sign for the two irreducible flavors,
then Euclidean correlators could violate positivity constraints.}, i.e. that
$D$ in (\ref{diracop3}) is replaced by
 \begin{equation}
 D=\begin{pmatrix}
 \slashed{D}+m&0\\ 0&\slashed{D}-m
 \end{pmatrix},
 \quad \slashed{D}=
 \gamma^\mu(\partial_\mu-\ii v_\mu)\,.
\label{diracop2}
 \end{equation}
The effective action in (\ref{eff_action1}) takes the form
\begin{equation}
S_\mathrm{eff}(v)=\lamth\int\D^3x\, v^\mu v_\mu-N\,\ln 
\det\big(m^2-\slashed{D}^2\big)\,.\label{eff_action2}
\end{equation}
As order parameter for chiral symmetry 
we use the chiral condensate
\begin{equation}
\Sigma=\frac{\ii}{2N}\sum_a\langle\bar{\psi}_a\Sigma_{45} \psi_a\rangle 
\,,\label{condensate1}
\end{equation}
where the insertion $\Sigma_{45}$ originates from
the change of variables in (\ref{changevar1}).
Using translational invariance it can be written as
\begin{equation}
\Sigma=
\frac{1}{V}\frac{1}{Z}\int\!\mathcal{D}v^{\mu}\;
\tr\left(\frac{m}{m^2-\slashed{D}^2}\right)e^{-S_\mathrm{eff}(v)}
\,.
\label{condensate2}
\end{equation}
We see here that only the Dirac operator $\slashed{D}$ of
\emph{one} irreducible flavor -- 
introduced in (\ref{diracop3}) -- enters
the expression for the partition function and 
chiral condensate of $N$ reducible flavors.
Note that the condensate is 
real and positive, $\Sigma=\vert\Sigma\vert$. 
In terms of the spectral density $\rho_v$ of
the irreducible Dirac-operator in a fixed auxiliary 
field, defined by 
\begin{equation}
\tr f(\ii\slashed{D})=\int_{-\infty}^\infty \D E\,f(E)
\rho_v(E)\,,\label{spectraldensity}
\end{equation}
the condensate (\ref{condensate2}) can be written as
\begin{equation}
\Sigma
=\frac {2m}{V}\int_0^\infty
\frac{\D E}{E^2+m^2}\,\bar\rho(E)\,,\label{condensate3}
\end{equation}
where the non-negative expectation value $\bar\rho(E)$
is calculated with the effective action,
\begin{equation}
\bar\rho(E)=\frac{1}{Z}
\int\! \mathcal{D}v^\mu\, e^{-S_\mathrm{eff}(v)}\,
\rho_v(E)=\bar\rho(-E)
\,.\label{av_density}
\end{equation}
The last relation follows from charge  conjugation 
symmetry which implies $\rho_v=\rho_{-v}$ and
is explained in appendix A.
In the limit $m\to 0$ equation (\ref{condensate3}) 
gives rise to a variant of the celebrated
Banks-Casher relation \cite{Banks:1979yr}.
It relates the low end of the spectral density
of the irreducible operator $\ii\slashed{D}$ 
to the chiral condensate of the reducible models.
In passing we note that --
because of parity-symmetry -- the would-be order parameter
of parity $\propto \langle\bar{\psi}_a\psi_a\rangle$
is identically zero for all reducible models.
This means that there is no spontaneous breaking of parity.

Finally, we must emphasize that varying 
the number of reducible flavors $N$ continuously
between $\frac{1}{2}$ and $1$ as described above
is not equivalent to varying the number of 
irreducible flavors $\Ni$ between $1$ and $2$.
There are several reasons for this
difference: First and 
rather technically, the $\Ni=1$ model
suffers from a severe sign problem and can only be simulated in an interesting dual formulation
\cite{Wellegehausen:2017goy}, in contrast to
the reducible
model with $N=\frac{1}{2}$, which has no
sign problem. Second and more important, 
for $\Ni=1$ the $\Z_{2}$ parity symmetry 
can be broken (by the anomaly and/or spontaneously) 
while parity is never broken
for the reducible systems. 

In the Thirring models with $\Ni=2$ 
and $N=1$ the global U$(2)$ chiral symmetry can be broken 
to U$(1)\times $U$(1)$ in which case we should
see two massless Goldstone bosons in
the particle spectrum. Finally we note, that the interpolating models 
with $N\notin \N/2$ probably do not describe 
local quantum field theories. But this problem 
will not invalidate the reasoning in the present work.

\section{Chiral condensate}\label{chap:condensate}
\noindent
We performed simulations with chiral 
SLAC-fermions on lattices $L \times (L-1)^2$ 
in the range $L=6\dots 24$. To control and
stabilize our simulations, we chose a mass proportional 
to the inverse lattice size, 
\begin{equation}
m=\frac{m_0}{L}\,,
\end{equation}
with small dimensionless parameter $m_0$. Note that for 
any fixed value of $m_0$ one recovers the 
massless Thirring model in the infinite volume 
limit $L \to\infty$.

\begin{figure}[tb]
 \scalebox{1.3}{\inputFig{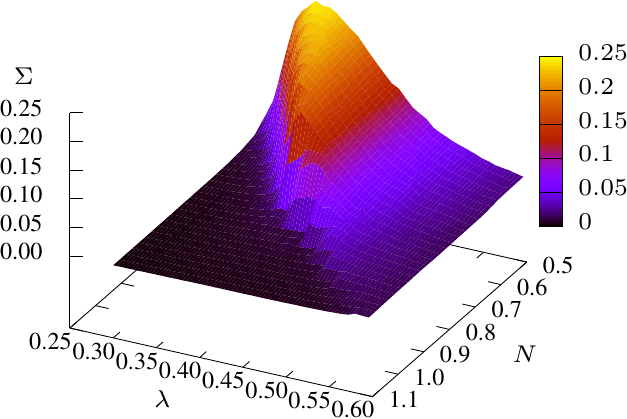}}
\caption{Expectation value of the chiral condensate 
$\Sigma$ as function of the coupling $\lambda$ and 
the number $\Nr$ of $4$-component spinors on 
a $16\times 15^2$ lattice with $m_0=0.1$.\label{Fig:Overview}}
\end{figure}

Figure~\ref{Fig:Overview} shows the surface plot of the
chiral condensate for $\lambda=0.25 \dots 0.60$ and $\Nr=0.5
\dots 1.1$ on a $16 \times 15^2$ lattice with $m_0=0.1$.
For $\Nr=0.5$ there exists a broken phase with
non-zero chiral condensate $\Sigma$. With increasing flavor number the chiral condensate falls off very quickly.
For small $\lamth$ the condensate
vanishes due to the (annoying but well-known) large lattice artifacts in the strong coupling regime,
\cite{DelDebbio1997,Wellegehausen:2017goy}.

In order to determine the critical flavor number, we
investigate the maximum $\Sigmam$ of the 
$\lambda$-dependent chiral condensate 
$\Sigma$ for different flavor numbers $N$ and
lattice sizes $L$. The maximum of the condensate is 
well-motivated since it clearly signals the
breaking of chiral symmetry.
The obtained results fully comply with those
obtained with the alternative method based on the 
susceptibility of the interaction term in a later
section.

\begin{figure}[tb]
 \scalebox{1.1}{\inputFig{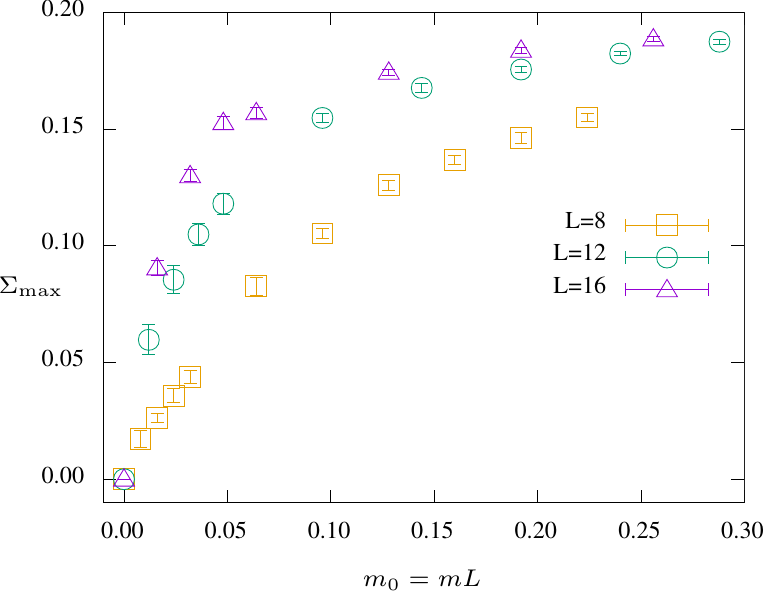}}
\caption{Maximal chiral condensate $\Sigmam$  
for different lattices volumes and $\Nr=0.70$ as 
function of the mass-parameter $m_0=m L$.\label{Fig:Mass}}
\end{figure}

\begin{figure}[tb]
 \scalebox{1.1}{\inputFig{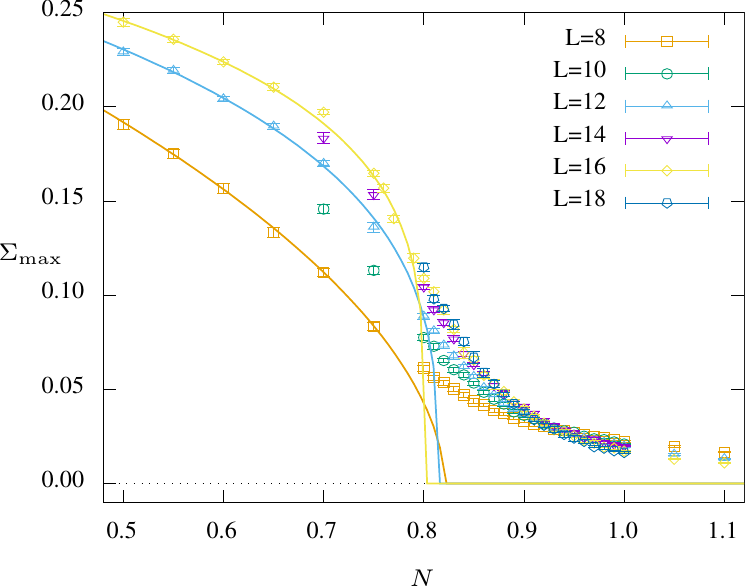}}
\caption{Maximal chiral condensate $\Sigmam$
	as function of $\Nr$ for $m_0=0.1$ on lattices with different sizes. Also shown are fits of the critical behavior according
	to Equation~\eqref{eq:critfit} and parameters from Table~\ref{tab:critfit}. \label{Fig:Maximum01}}
\end{figure}

Figure~\ref{Fig:Mass} shows the dependence of
$\Sigmam$ on the mass parameter $m_0$ for 
three different lattice volumes and for $\Nr=0.70$.
For a fixed $m_0$ (with Compton wave-length much
smaller than the lattice size) the chiral condensate increases 
with increasing lattice volume.
Performing the infinite volume limit -- which includes the
$m\to 0$ limit for every $m_0>0$ -- we conclude that
for $\Nr=0.7$ chiral symmetry is spontaneously broken.
Actually, in most of our simulations we choose $m_0=0.1$, 
which is a good compromise between good chiral properties,
simulation performance and small finite volume effects.
The results for the maximal condensate $\Sigmam$ as function
of $\Nr$ (for $m_0=0.1$) is depicted in Figure~\ref{Fig:Maximum01}.
For a fixed lattice volume, the 
condensate increases with decreasing flavor number. 
For a fixed $\Nr\lessapprox 0.8$ the maximal condensate 
increases with increasing lattice volume and we conclude
that chiral symmetry is broken for these $\Nr$.
We compared with the results obtained with $m_0=0.04$
and obtained a comparable outcome. But for this smaller mass
finite size effects are less suppressed.
\begin{figure}[tb]
 \scalebox{1.1}{\inputFig{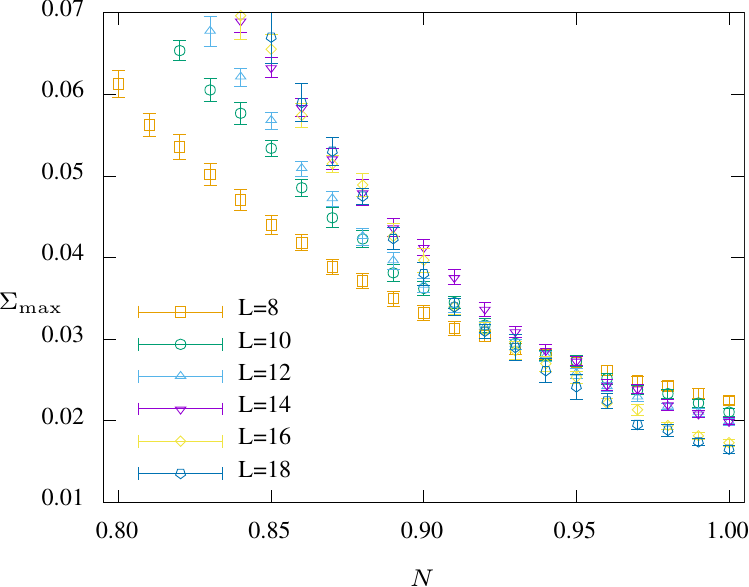}}
\caption{Maximal chiral condensate 
	$\Sigmam$ as function of $\Nr$ 
	above the critical flavor number and
	for $m_0=0.1$ on lattices with different sizes.\label{Fig:Maximum01Detail}}
\end{figure}
\begin{figure}[tb]
	\scalebox{1.1}{\inputFig{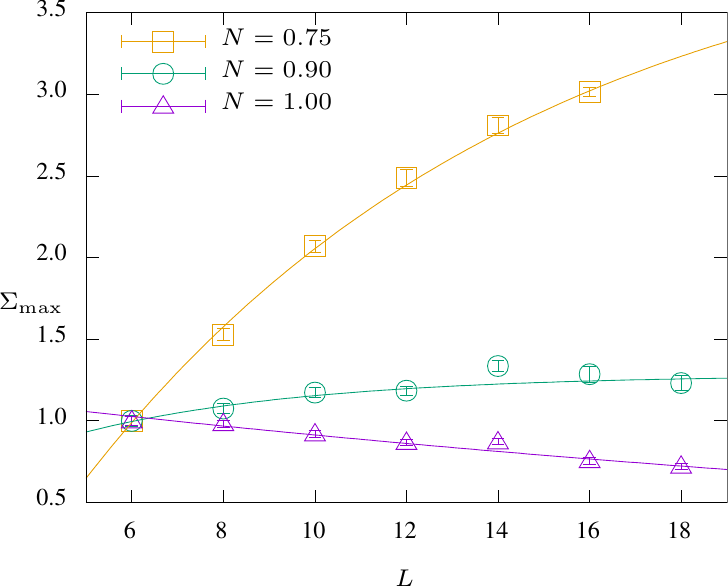}}
	\caption{Maximum of the condensate (normalized to the smallest lattice) as function of the lattice size $L$ for three different values of $\Nr$.\label{Fig:Maximum01Volume}}
\end{figure}
The region above $\Nr=0.8$ is magnified in
Figure~\ref{Fig:Maximum01Detail}. Above
$\Nr=0.95$ the condensate decreases with increasing
volume and one concludes that chiral symmetry remains unbroken 
in this regime. Unfortunately, the lattices are not 
sufficiently large to permit a reliable extrapolation to 
infinite volume for all values  of $\Nr$ under consideration. 
That was only achieved for the flavor numbers
below $0.75$ and above $1.00$. Three examples are depicted in Figure~\ref{Fig:Maximum01Volume}. Since we introduced a
mass, we expect a finite size scaling law of the form \cite{neuberger_finite_1989}
\begin{align}
	\label{eq:fit}
	\Sigmam(L) = ae^{-bL}+\Sigmam(\infty)
\end{align}
for which the optimal fit-parameters in
the fits depicted in Figure~\ref{Fig:Maximum01Volume} are 
listed in Table~\ref{tab:fit}.\footnote{In all the following fits, $L=14$ was excluded from fitting because there were 
some incurable problems with the thermalization. Also,
$\Sigma(\infty)$ was constrained to be positive, since
the exact condensate has this property.}
\begin{table}[h!]
	\caption{Fit parameter for the infinite volume extrapolation of the maximal chiral condensate for three values of $\Nr$.\label{tab:fit}}\vskip1mm
	\begin{tabular}{|c|c|c|c|}
		\hline
		$\Nr$ & $a$ & $b$ & $\Sigmam(\infty)$\\
		\hline
		$0.75$ & $-5.9(1)$ & $0.10(2)$ & $4.21(4)$\\
		$0.9$ & $-1.0(3)$ & $0.20(9)$ & $1.29(6)$\\
		$1.0$ & $1(1)$ & $0.03(4)$ & $0(1)$\\
		\hline
	\end{tabular}
\end{table} 
In the broken phase with small $\Nr$ (e.g.\@ $0.75$)  this
extrapolation works well. Also for \mbox{$\Nr=1.0$} the
exponential function (\ref{eq:fit}) fits the data 
well and points to a
vanishing condensate in the infinite volume limit. 
For values of $\Nr$ in between the data becomes basically flat
due to large finite size effects -- in some cases they are
even non-monotonic -- which renders an extrapolation
unreliable. 

However, for every finite volume we find that the maximal 
chiral condensate exhibits a turning point around $\Nr\approx 0.8$
where the chiral condensate is bending upwards, see Figure~\ref{Fig:Maximum01}. This bending is caused
by finite size effects and the explicit breaking of chiral
symmetry by the fermion mass term. The data points to the left 
of this turning point are well-described by
the scaling law
\begin{align}
\label{eq:critfit}
\Sigmam(\Nr)=a(\Nrc-\Nr)^{\beta}
\end{align}
with parameters $a,\Nrc,\beta$ given in Table~\ref{tab:critfit}.
In particular we can read off the critical
flavor number and conclude, that there is no 
spontaneous symmetry breaking above
\begin{align}
\label{eq:Nrc}
	\Nrc = 0.80(4)\,.
\end{align}
In the following sections we will  substantiate the
result (\ref{eq:Nrc}) with other methods.
Note that our lattice volumes are not large enough to extract
a reliable value for the critical exponent $\beta$.
But since our main focus is on the critical flavor number,
which does not suffer from finite size effects, we did
not further increase the lattice volume to
obtain a more accurate value for $\beta$.
\begin{table}
	\caption{Fit parameters for fitting the critical behavior.\label{tab:critfit}}\vskip1mm
	\begin{tabular}{|c|c|c|c|c|}
		\hline
		$L$ & $\Nr$ & $a$ & $\beta$ & $\Nrc(L)$\\
		\hline
		$8$ & $[0.50,0.75]$ & $0.245(4)$ & $0.6(1)$ & $0.82(4)$\\
		$12$ & $[0.50,0.80]$ & $0.266(3)$ & $0.31(2)$ & $0.815(3)$\\
		$16$ & $[0.50,0.79]$ & $0.276(4)$ & $0.23(2)$ & $0.801(4)$\\
		\hline
	\end{tabular}
\end{table}
The critical exponent $\beta$ has been calculated previously
with the functional renormalization group
(FRG), with Dyson-Schwinger equations (DSE) and with
Monte-Carlo simulation with staggered fermions (MC).
We compiled some results 
with references in Table~\ref{tab:beta}.
\begin{table}
	\caption{\label{tab:beta}Compilation of various
		results for the
	critical exponent $\beta$ from
	the literature. The abbreviations are explained in the main
	text.}\vskip1mm
	\begin{tabular}{|c|c|c|}
		\hline
		Method&$\beta$&Ref.\\
		\hline
		FRG&0.44&\cite{Janssen2012a}\\
	 	DSE&1&\cite{Kondo1995}\\
		MC (staggered)&0.37&\cite{Christofi2007}\\
		\hline
	\end{tabular}
\end{table}
We see that the predictions for the critical exponent
$\beta$ depend much on the non-perturbative method in use.
But the quoted values cannot be easily compared among themselves
and with our results in Table \ref{tab:critfit}. 
For example, with staggered fermions one may
simulate another universality class. We intend to find 
a better value of $\beta$ with chiral fermions
on  larger lattices in the future.

For the smaller mass parameter $m_0=0.04$ we obtain 
qualitatively the same data. However, the ill-conditioned 
fermion determinant forbids a more detailed study for 
this (and smaller) masses.

\section{Spectral density}\label{chap:density}
\begin{figure}[tb]
\scalebox{1.1}{\inputFig{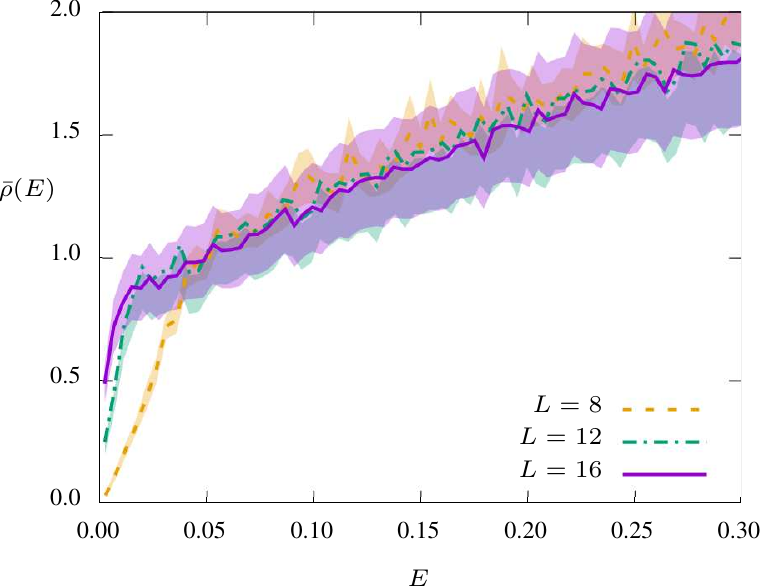}}
\caption{Mean spectral density 
		$\bar\rho(E)$ in (\ref{av_density})
		for $\Nr=0.80$ (brooken phase) for different lattice sizes.
		The shaded regions indicate the uncertainties.
		\label{Fig:Eigenvalues160}}
\end{figure}

\noindent
As explained above, the chiral properties of the theory 
can be extracted from the spectral density $\rho_v(E)$ 
of the massless irreducible Dirac operator
introduced in (\ref{spectraldensity})
and the average spectral density $\bar{\rho}(E)$ 
defined in (\ref{av_density}).
If $\bar\rho(E)$ in the neighborhood of $E=0$
remains small with increasing volume, then
chiral symmetry is realized.
On the contrary, if it increases,
then chiral symmetry is broken. 
Figure~\ref{Fig:Eigenvalues160} shows the spectral
density for $N=0.80$ on different lattice sizes.
Close to the origin, the density clearly builds up with 
increasing lattice volume and one concludes that 
chiral symmetry is broken.
For the larger flavor number $\Nr=1.00$ we observe the
opposite behavior, see Figure~\ref{Fig:Eigenvalues200}:
\begin{figure}[tb]
 \scalebox{1.1}{\inputFig{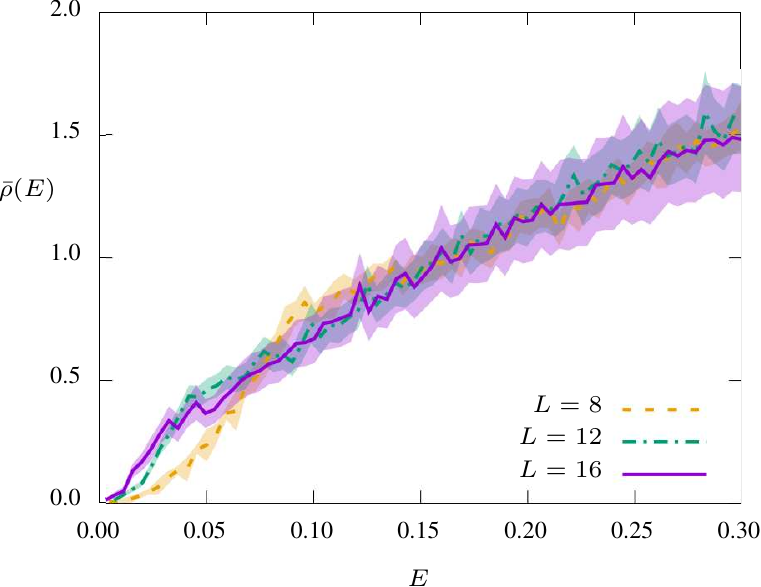}}
\caption{Mean spectral density 
	$\bar\rho(E)$ in (\ref{av_density})
	for $\Nr=1.00$ (symmetric phase) for different lattice sizes. The shaded regions indicate the uncertainties.
	\label{Fig:Eigenvalues200}}
\end{figure}
Close to the origin, the density remains small for all
lattice sizes. Again we conclude that for
$\Nr=1.00$ chiral symmetry is unbroken.

\section{Goldstone spectrum}\label{chap:spectrum}
\noindent
Next we investigate the meson spectrum of the 
$N$-flavor theory. There are two scalar and two
pseudoscalar mesons with vanishing angular momentum
and the corresponding interpolating operators are $\cO_\Gamma=\frac{1}{N}\sum_a\bar{\psi}_a\Gamma\psi_a$,
where $\Gamma$ is the identity matrix or one of the
three matrices $\ii\Gamma_4,\ii\Gamma_5$ and
$\Gamma_{45}$ in (\ref{gammas}). Since all reducible
flavors contribute equally to $\cO_\Gamma$, we may set $N=1$ 
in these bilinears. Thus we choose the operator basis
\begin{equation} 
\cO_a(t)=\sum_{\vx}\cO_a(t,\vx),\quad a=0,1,2,3\,,
\end{equation}
which are the zero-momentum projections of
\begin{equation}
\cO_a(x)=\bar{\psi}(x)(\sigma_a\otimes\sigma_0)
\psi(x)=\cO_a(t,\vx)\,,
\end{equation}
and where $\psi$ represents one of the $N$ reducible flavors.
For example, $\sigma_1\otimes\sigma_0$ swaps the
two irreducible spinors which make up the
reducible $4$-component spinor.
Note that the expectation value of $\cO_3(x)$ 
is twice the chiral condensate.
In our simulations, we measure the 
correlation matrix with elements
\begin{align}
C_{ab}(t)&=\vev{\cO_a(t)\cO_b(0)}-\vev{\cO_a(t)}
\vev{\cO_b(0)}\nonumber \\ 
&=\sum_{\vx,\vy}
\big\langle\tr{\sigma_a\Delta_{xx}}\tr{\sigma_b\Delta_{yy}}
-\tr(\sigma_a\Delta_{xy}\sigma_b\Delta_{yx})\big\rangle\nonumber\\
&\qquad -\,\sum_{\vx,\vy}
\vev{\tr\sigma_a\Delta_{xx}}\vev{\tr\sigma_b\Delta_{yy}}\,,\label{corrmatrix}
\end{align}
where $\Delta$ is the propagator for $4$-component
fermions in a fixed auxiliary field $v_\mu$,
\begin{equation}
\Delta=\frac{1}{D}=\begin{pmatrix} \Delta_+ & 0 \\ 0 & \Delta_-\end{pmatrix},\quad 
\Delta_{\pm}=\langle x\vert\frac{1}{\ii\slashed{D}\pm\ii m}\vert y\rangle\,.
\end{equation}
The expectation values in (\ref{corrmatrix}) are calculated
with $S_\mathrm{eff}$ and traces are taken in spinor and 
flavor space. By exploiting parity invariance we prove in 
appendix A that the correlation matrix is diagonal.
It is most conveniently expressed in terms of the
parity odd and parity even terms in the decomposition
\begin{equation}
\Delta=\sigma_0\otimes A+\sigma_3\otimes B\,,
\label{decomp_green1}
\end{equation}
where
\begin{equation}
A
= \frac{\ii\slashed{D}}{m^2-\slashed{D}^2}\;,\quad
\ii B=
\frac{m}{m^2-\slashed{D}^2}\,.\label{decomp_green2}
\end{equation}
The diagonal elements of $(C_{ab})$ -- these are the 
eigenvalues -- read
\begin{align}
\begin{split}
C_0(t)&=
4\sum_{\vx,\vy}\big\langle\trd A_{xx}\trd A_{yy}\big\rangle
\\
&-2\sum_{\vx,\vy}\big\langle
\trd(A_{xy}A_{yx}+ B_{xy}B_{yx})\big\rangle\,,
\\
C_1(t)&=\;C_2(t)=2\sum_{\vx,\vy}\big\langle 
\trd(B_{xy}B_{yx}- A_{xy}A_{yx})\big\rangle\,,
\\
C_3(t)&=
4\sum_{\vx,\vy}\big(\big\langle\trd B_{xx}\trd B_{yy}\big\rangle
+\Sigma^2\big)
\\
&-2\sum_{\vx,\vy}\big\langle 
\trd(A_{xy}A_{yx}+B_{xy}B_{yx})\big\rangle\,,\label{corr_matrix1}
\end{split}
\end{align}
where $x=(t,\vx)$ and $y=(0,\vy)$.

If chiral symmetry is spontaneously broken according to
\begin{equation} 
\text{U}(2)
\to \text{U}(1) \otimes \text{U}(1)
\end{equation}
two of the four mesons should become massless Goldstone bosons while 
the other two remain massive. More precisely, since the 
interpolating operators $\cO_1$ and $\cO_2$ correspond to 
the Goldstone bosons, their correlators $C_{1}=C_{11}$ and $C_{2}=C_{22}$ should describe massless particles.

\begin{figure}[tb]
 \scalebox{1.1}{\inputFig{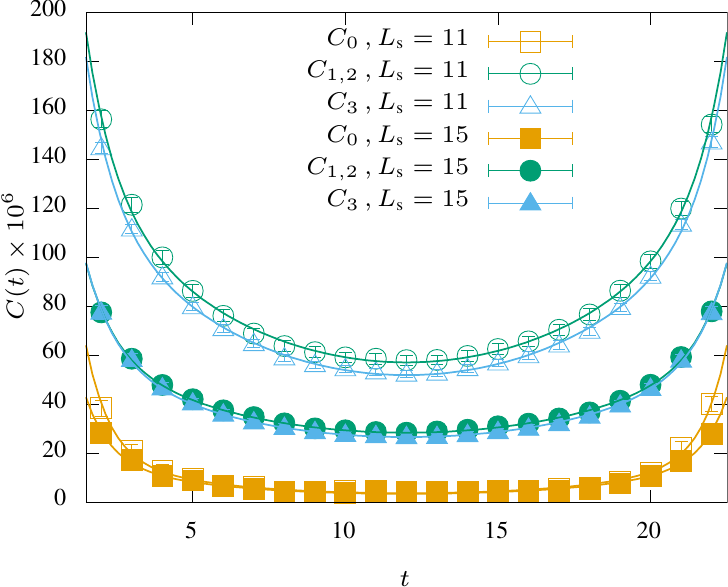}}
\caption{Meson correlation functions for $\Nr=1.00$ (symmetric
	phase) on a $L_\text{s}^2 \times 24$ lattice and $m=0.004$.}\label{Spectrum_s}
\end{figure}

If chiral symmetry is not broken, we expect that the 
four (pseudo)scalars 
arrange in a singlet and a triplet of SU$(2)\subset$
U$(2)$.
In particular, the state belonging to the interpolating
operators $\cO_1,\cO_2$ and $\cO_3$ should form a triplet.
In the corresponding subspace the correlation matrix has
eigenvalues $C_{1},C_2$ and $C_3$.
Indeed, in the symmetric phase we have $B_{xy}=0$ for
$m\to 0$ and these $3$ eigenvalues become
identical,
\begin{align}
C_0(t)&=\sum_{\vx,\vy}\big\langle 4\trd A_{xx}\trd A_{yy}
-2\trd(A_{xy}A_{yx})\big\rangle\,,\nonumber\\
C_1(t)&=C_2(t)=C_3(t)=-\sum_{\vx,\vy}\big\langle 2\trd(A_{xy}A_{yx})\big\rangle\,.
\end{align}
In Figure~\ref{Spectrum_s} we show the (pseudo)scalar spectrum 
in the symmetric phase at $\Nr=1.00$ for two different lattice 
volumes $11 \times 24$ and $15 \times 24$ and a residual 
mass $m=0.004$.
The correlation functions $C_1,C_2$ and $C_3$ for 
both spatial volumes lie almost on top of each other --
the splitting originates from the explicit breaking by
the mass term --
while $C_0$ decays faster. The lines represent fits with 
a sum of two cosh-functions for the ground and excited state. 
The fitted masses are given in Table \ref{tab:fitMass}. 
For both the ground and excited multiplet we find three 
almost identical masses and a larger one. Within 
statistical uncertainties and taking into account
finite volume effects, the results are compatible with 
two multiplets of massive mesons in the symmetric phase.
\begin{figure}[tb]
 \scalebox{1.1}{\inputFig{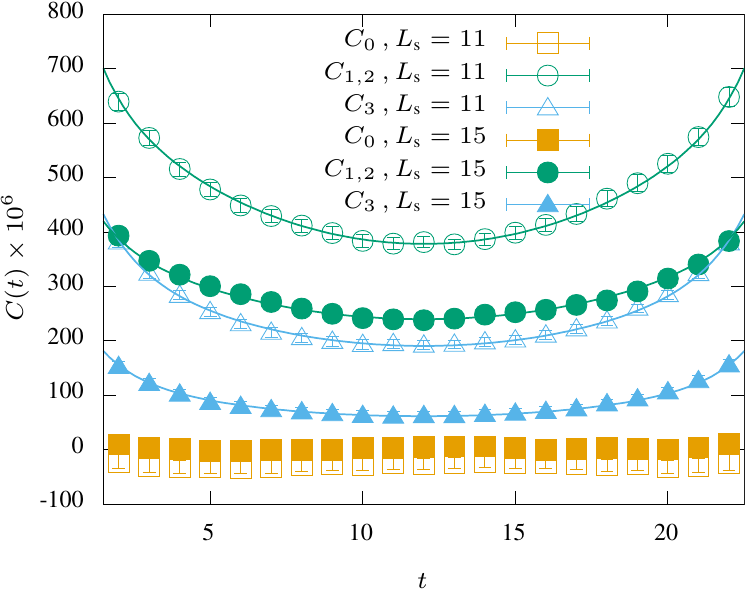}}
\caption{Meson correlation functions for $\Nr=0.80$ 
	(broken phase) on a $L_\text{s}^2 \times 24$ lattice 
	and $m=0.004$}.\label{Spectrum_b}
\end{figure}

In the broken phase at $\Nr=0.80$, see Figure~\ref{Spectrum_b}, 
the correlation functions $C_1=C_{2}$ and $C_3$ differ significantly
compared to the correlators in the 
symmetric phase. While the masses of $C_1=C_2$ are almost volume
independent, the ground state mass of $C_3$ shows stronger finite volume effects. 
The correlation
function $C_0$ is compatible with zero for all temporal extents $t$
which is explained by a corresponding correlation length not much bigger than
the lattice constant.
Thus the simulation results are compatible with the existence 
of two massless Goldstone bosons and two massive excited states 
with different masses. We conclude,
that the chiral U$(2)$ symmetry is indeed broken to U$(1) \otimes
\text{U}(1)$ for $\Nr=0.80$.
\begin{table}[tb]
	\caption{Meson masses in the symmetric phase
		with $\Nr=1.00$ and the broken phase with $\Nr=0.80$
		for two different spatial lattices $L_\text{s}=11$ and $L_\text{s}=15$.\label{tab:fitMass}}
 \begin{tabular}{|c|c|c|c|c|c|}
 \hline
$C$ & $m(11)$ & $m(15)$ & $m^*(11)$ & $m^*(15)$ & $\Nr$\\
\hline
$C_0$ & $0.21(2)$ & $0.21(2)$ & $1.27(6)$ & $1.22(7)$ & $1.00$\\
$C_{1,2}$ & $0.134(3)$ & $0.128(2)$ & $1.03(5)$ & $1.02(3)$ & $1.00$\\
$C_3$ & $0.138(2)$ & $0.131(2)$ & $1.08(4)$ & $0.98(3)$ & $1.00$\\
\hline\hline
$C_0$ & $-$ & $-$ & $-$ & $-$ & $0.80$\\
$C_{1,2}$ & $0.103(2)$ & $0.095(3)$ & $1.04(12)$ & $0.93(17)$ & $0.80$\\
$C_3$ & $0.109(4)$ & $0.127(7)$ & $0.81(7)$ & $0.81(10)$ & $0.80$\\
\hline
 \end{tabular}
\end{table}

In an interesting recent work by Simon Hands  
with bulk domain wall fermions 
(DWF) on a $12^3$-lattice 
(and $L_s$ up to $40$) the meson correlators of 
the $\Nr=1$ model have been calculated as well
\cite{Hands:2018vrd}.
Whereas an earlier simulation with
surface DWF on a
$12^2\times 24$ lattice (but $L_s$ only up to $16$) 
showed no sign of a chiral phase transition
for $\Nr=2$ \cite{Hands:2016foa}, 
the new results for $\Nr=1$ with DWF signal a Goldstone
spectrum expected from a U(2)$\to$ U(1)$\,\otimes$ U(1)
breaking. This means that for $1$ flavor 
the prediction of DWF are in conflict with our findings.
\section{Susceptibility}\label{sec:susc}
If for $\Nr\geq\Nrc$ the lattice artifact phase
transition were the only phase transition then one could hardly
imagine how to construct
an interacting QFT in the continuum limit.
And there are convincing arguments based on different
approaches that there exists a well-defined continuum
limit, corresponding to a UV-stable fixed point of the 
renormalization group (RG) \cite{Parisi1975,
Hikami:1976at,Gomes1991,Gawedzki:1985ed,Gies2010}.
To find the continuum theory at the transition to the artifact 
phase at strong bare couplings -- see \cite{Wellegehausen:2017goy}
for details -- seems unlikely since this transition only exists 
in a discretized setup. However, already in the quoted
work we have spotted signals of 
another transition in the intermediate coupling regime.
In this section we will argue, that such a transition
indeed exists for $\Nr\geq \Nrc$ and probably is 
continuous. In our earlier work we did not further
analyze this feature, mainly since scanning the
phase diagram of a fermion theory on lattices
of different sizes is rather expensive. For the same
reason we do not aim at a detailed finite size analysis
in the present work. But we do simulations on lattices
with different sizes to see the qualitative behavior
of the susceptibility related to the four-Fermi term
in the Lagrangian. Actually, the similarly accurate 
number for $\Nrc$ is extracted by spotting the 
merging of the newly discovered  
transition with the lattice artifact 
transition \footnote{We cannot be sure that the two transitions
	meet. But we will see that they come close.}.

\begin{figure}[tb]
	\scalebox{1.3}{\inputFig{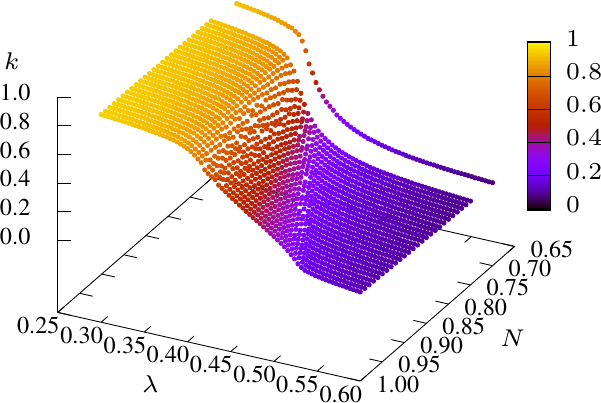}}
	\caption{Lattice filling factor $k$ as function of $\lambda$ and $\Nr$ on a
		$16\times 15^2$ lattice with $m_0=0.1$.\label{Fig:k}}
\end{figure}
\begin{figure}[tb]
  \scalebox{1.1}{\inputFig{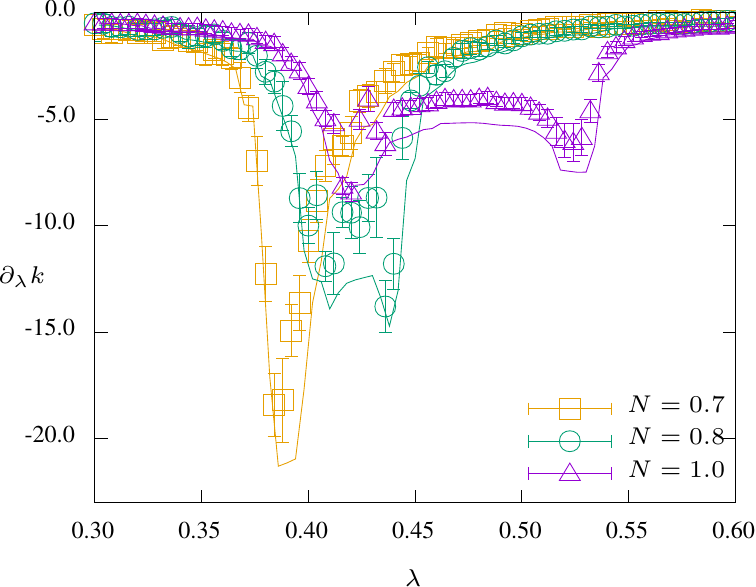}}
  \caption{Numerical derivatives for 3 different flavor numbers
  	$\Nr$.
  Markers denote a local derivative stencil and lines a global
  differentiation scheme (see appendix B).}\label{Fig:dk}
 \end{figure}

As tracer for the transition we
will consider the second derivative of the partition function
with respect to the coupling $\lambda$. As discussed in detail 
in our previous paper \cite{Wellegehausen:2017goy}, the 
partition function's first derivative can be associated 
with the lattice filling factor $k$ as follows,
\begin{equation}
	k = -\frac{\lambda}{\Nr V}\frac{\partial\ln Z(\lambda)}{\partial\lambda} + \mathrm{c}
	=\frac{1}{4N\lambda}\,\big\langle j^\mu j_\mu\big\rangle
	+ \mathrm{c}\,.
\end{equation}
and thus is (up to a $\lambda$-independent additive constant
c) proportional
to the expectation value of the four-Fermi 
interaction term $(\bar{\psi}\gamma^\mu\psi)^2$.
Roughly speaking, $k$ is the average fraction of 
lattice sites at which an interaction takes place. 
From this interpretation the following properties 
(established in \cite{Wellegehausen:2017goy}) are 
comprehensible: The filling factor vanishes
in the weak coupling regime (large $\lambda$),
it monotonically approaches $1$ when approaching the lattice
artifact phase at strong coupling and its derivative 
\begin{align}
\partial_\lambda k = 
-\frac{1}{16N\lambda^3}\sum_x \big\langle (j^\mu j_\mu)(x)
(j^\mu j_\mu)(0)\big\rangle_c
+\frac{\mathrm{c}-k}{\lambda}\label{suscept3}
\end{align}
exhibits a dip at this transition, since the
susceptibility of $j^\mu j_\mu$ -- this is just the sum 
over the first argument of the connected
two-point function on the right hand side -- 
peaks at the transition. 
All of these features are 
clearly seen in Figure~\ref{Fig:k}
and Figure~\ref{Fig:dk}.

What has not been discussed before is the fact that for 
not too small $\Nr$ the derivative $\partial_{\lambda}k$
shows  a second dip corresponding to a second
peak of the susceptibility of the $4$-Fermi 
term at intermediate values
of $\lambda$. This was already visible
at the edge of Figure 4 in \cite{Wellegehausen:2017goy}.
Since the direct computation of $\partial_\lambda k$ as
$8$-point function would be rather expensive computationally, 
we instead use the numerical derivative of $k$ to 
calculate the susceptibility. But conventional 
finite-difference approximations of the $\lambda$-derivative
will greatly amplify the noise present in our data. 
There are many methods to regularize the differentiation 
process (regression, smoothing, filtering, variation denoising).
In our analysis we used a variation denoising method (and 
compared it with the conventional approach).
More details can be found in
appendix B.

\begin{figure}[tb]
	\scalebox{1.1}{\inputFig{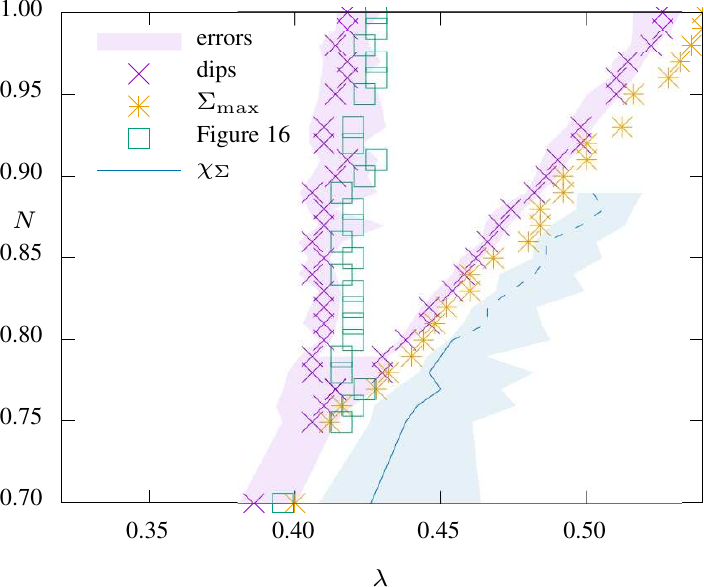}}
	\caption{Extracted dip positions of $\partial_\lambda k$ supplemented with the locations where $\Sigma$ assumes its maximum (marked by $\Sigma_\mathrm{max}$)
		and the positions of evaluation for Figure~\ref{Fig:cc2}. The maximum of the 	chiral susceptibility 
		(marked by $\chi_\Sigma$) which
signals the \emph{chiral phase transition} is 
shown as solid line.}\label{Fig:peaks}
\end{figure}

Examples of $\partial_\lambda k$ 
at three different flavour numbers $\Nr$ are depicted 
in Figure~\ref{Fig:dk}. 
One recognizes two qualitatively different
behaviors. For small $\Nr$ ($\Nr=0.7$ in Figure~\ref{Fig:dk})
$\partial_\lambda k$ has one distinct minimum which -- 
as discussed before -- signals the transition into the 
lattice artifact
phase. For sufficiently large $\Nr$ ($\Nr=1.0$ in 
Figure~\ref{Fig:dk}) two distinct minima are clearly visible. 
A comparison with the data presented in Figure~\ref{Fig:k}
reveals that the minimum to the left (strong coupling)
signals the transition into the lattice artifact phase.
The second (inverted) peak at intermediate coupling has not been
discussed before and the following discussion makes
clear that it belongs to an interaction driven transition
at intermediate couplings.
For flavor numbers near
\mbox{$\Nr=0.8$} the two minima merge and the artifact
phase transition line comes close to or meets the line of
interaction driven-phase transitions.

The numerically extracted peak
positions of the susceptibility for $\Nr$ 
between $0.7$ and $1.0$ are shown in Figure~\ref{Fig:peaks}.
At first glance one can see that for $\Nr<0.76$
there is only one phase transition and for $\Nr>0.82$
there are two transitions.
More accurately, on a lattice with $L=16$ the two 
transition lines get close or
meet at the triple point at
\begin{align}
	N^\mathrm{t}(L=16) = 0.78(4).
\end{align}
This value matches the putative critical flavor 
number $\Nrc$ in \eqref{eq:Nrc} pretty well. Our
explanation of this only seemingly surprising
equality is the following: 
the ubiquitous lattice artifact phase at strong coupling
does not describe any properties of the continuum 
Thirring model. Only in the physical phase
at weaker couplings can we hope to construct a continuum
theory when approaching a critical point or critical line of
second order transitions. For
sufficiently small $\Nr$ the perfect candidate for this
transition is the chirality breaking transition discussed 
previously. Indeed, the line where the 
condensate is maximal is always to the right of 
\emph{both} minima of $\partial_{\lambda}k$,
see Fig. \ref{Fig:peaks}.
Hence for $\Nr\geq N^\mathrm{t}$ the maximum
is to the right of the interaction driven phase transition
line and we have seen that this maximal
condensate decreases with increasing lattice size.
The only plausible explanation is the following: for 
small $N\leq \Nrc$ there is, beside
the artifact transition at strong coupling a 
second order chiral phase transition at intermediate 
coupling. The solid line in Fig.~\ref{Fig:peaks} indicates 
the position of the maximal chiral susceptibility (obtained as numerical derivative $\chi_\Sigma = -\partial_{\lambda}\Sigma$) 
where the chiral phase transition on sufficiently large
lattices happens. At $\lambda\approx 0.42$ and $\Nr=\Nrc$ 
the chiral phase transition line comes
close to the artifact transition line and ceases to exist
since in the thermodynamic limit there is no non-vanishing
 condensate for $\Nr\geq \Nrc$. 
Instead a (probably) second order phase transition 
line emerges for $\Nr\geq N^\mathrm{t}\approx \Nrc$ where 
the derivative $\partial_{\lambda}k$ develops a
singularity. The second order chiral transition
with order parameter turns into a second order 
interaction-driven transition without order 
parameter.

\begin{figure}[tb]
	\scalebox{1.1}{\inputFig{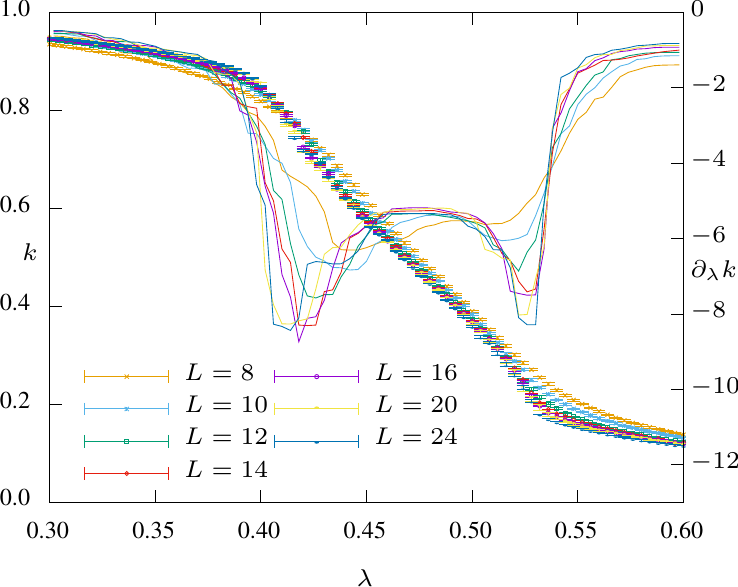}}
	\caption{Lattice filling factor $k$ (markers) and 
		its derivative $\partial_\lambda k$ (lines) 
		for various lattice sizes at $\Nr=1.0$.}\label{Fig:dkVol}
\end{figure}

\begin{figure}[tb]
	\scalebox{1.1}{\inputFig{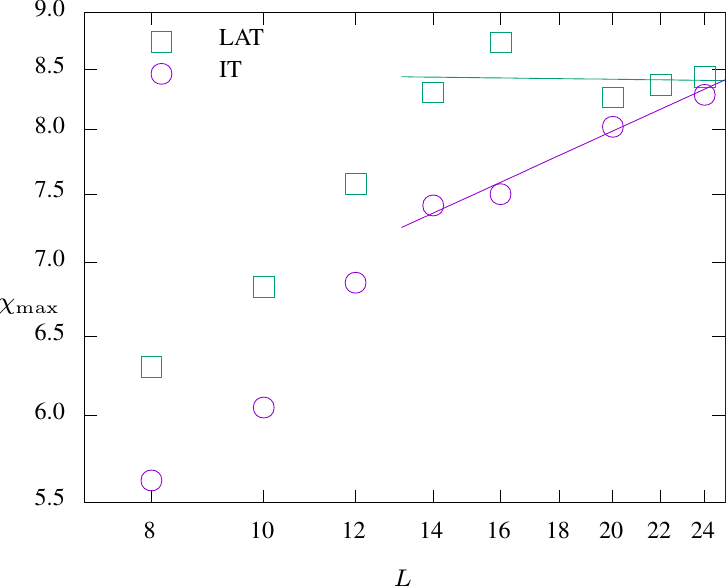}}
	\caption{Depth of the minima at the lattice artifact transition (LAT) and the interaction-driven transition (IT) at $\Nr=1.0$ on a log-log scale with fits according to Eq.~\eqref{eq:suscFit}.}\label{Fig:suscMax}
\end{figure}

\begin{figure}[tb]
	\scalebox{1.1}{\inputFig{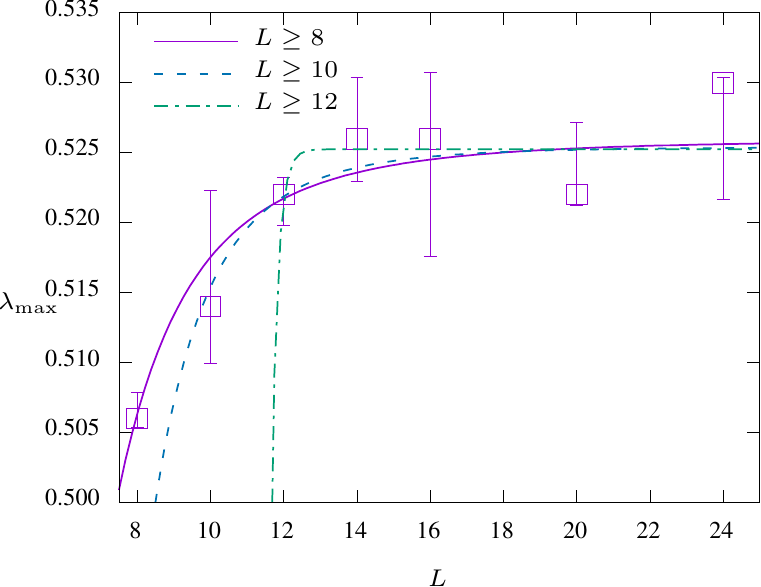}}
	\caption{Position of the interaction-driven transition at $\Nr=1.0$ with fits according to Eq.~\eqref{eq:lambdaCfit}.
	}\label{Fig:suscPos}
\end{figure}

The order of a phase transition
is related to the dependence of the peak susceptibility
on the size of the system. For $\Nr=1.0$ this 
behavior of $k$ and $\partial_{\lambda}k$ is depicted in
Fig.~\ref{Fig:dkVol} while the depth of the minima and the position
of the interaction-driven transition are shown in
Fig.~\ref{Fig:suscMax} and \ref{Fig:suscPos} respectively. In both
figures, one can see that finite size effects are significant for
$L<14$. Above that, at the 
artifact transition the susceptibility $\propto\partial_{\lambda}k$
is almost independent on the volume as expected for a first order
transition. On the other hand, at the interaction-driven
transition the susceptibility increases with increasing 
volume roughly according to \cite{fehske_monte_2008}
\begin{align}
	\label{eq:suscFit}
	\ln \chi_{\mathrm{max}}(L)=a\ln L + b
\end{align}
as expected 
for a second order transition, where 
$\chi\propto -\partial_{\lambda}k+\mathrm{const}$ is the susceptibility of the $4$-Fermi term,
see (\ref{suscept3}).
Such a linear dependence is actually seen in 
the double-logarithmic plot in Figure \ref{Fig:suscMax}
for lattices with $L\geq 14$. 
However, the data points for lattices with $L\geq 14$ 
are too noisy and we cannot extract reliable values for 
the fit parameters $a$ and $b$.
The theory of finite-size scaling also predicts
that the coupling $\lambda_{\mathrm{max}}(L)$, where
$\chi(L)$ peaks, approaches the critical coupling in the 
thermodynamic limit $\lambda_{\mathrm{c}}$ as \cite{fehske_monte_2008}
\begin{align}
	\label{eq:lambdaCfit}
	\lambda_{\mathrm{max}}(L)&=\lambda_{\mathrm{c}}(1-cL^{-1/\nu}).
\end{align}
We observe that at the interaction-driven 
transition this scaling law reproduces the
data reasonably well. The extracted value
\begin{align}
	\lambda_{\mathrm{c}}=0.526(5),
\end{align}
is rather stable -- it does not change much
if the fits are for lattice with
$L\geq 8,\,L\geq 10$ or $L\geq 12$, see
Figure \ref{Fig:suscPos}. But the extracted values
for $\nu$ vary considerably and cannot be
trusted. The lattices considered are just not big
enough to determine the critical exponent $\nu$.
But the aim of our very crude finite size analysis is not
to calculate critical exponents but rather to study
the order of the interaction-driven transition.
Our preliminary results suggest that it is a second order 
transition with infinite correlation length.

Most likely
this continuous transition is not associated with any
symmetry breaking, since the term $(\bar{\psi}\Gamma_{\mu}
\psi)^2$ is already part of the Thirring-Lagrangian \eqref{eq:lagrangian}. 
Such transitions without change of symmetry are common 
in condensed matter physics and they are sometimes 
called iso-symmetric. In solid state physics such 
transitions are structural
and are related to discontinuous changes of the cell 
volume and cell parameters and thus indicate a first-order 
transition. But continuous transitions without
symmetry breaking are also possible in which case
we prefer the name interaction-driven 
transitions. For example, a continuous transition without
symmetry-breaking bilinear fermion condensate -- triggered by
a four-fermion interaction term -- has been reported previously
in SU(4)-invariant four-fermi models 
in three dimensions. These models are similar to the Thirring 
model considered in the present work.
Numerical simulations with staggered fermions,
the fermion bag method, 
hybrid Monte Carlo and quantum Monte Carlo revealed 
actually an interesting phase structure 
\cite{Ayyar:2014eua,Catterall:2015zua,He:2016sbs,Ayyar:2016lxq}:
The systems exhibit
a continuous quantum phase transition from a
weakly coupled massless phase (a gapless Dirac semimetal)
to a massive (fully gapped Mott insulator) phase without 
condensing any fermion bilinear operator. It could very 
well be that a similar mechanism is at work in the 
Thirring models, although a bilinear condensate
is not forbidden by symmetry arguments as it is in
the SU(4)-invariant models.

Although the transition is probably not associated
with a change of symmetry there could exist an
order parameter. The filling factor $k$
is a possible candidate.
From Figure~\ref{Fig:dkVol} one might conjecture that 
for weak coupling (to the right of the interaction-driven transition)
$k$ approaches $0$ in the infinite volume limit.
This would imply that only the phase between the peaks 
describes an interacting four-Fermi theory. Actually, we can
prove that $k=1$ in the strong coupling 
expansion, see \cite{Wellegehausen:2017goy}, but
so far we could not show that $k=0$ in a weak 
coupling expansion.

\begin{figure}[tb]
	\scalebox{1.1}{\inputFig{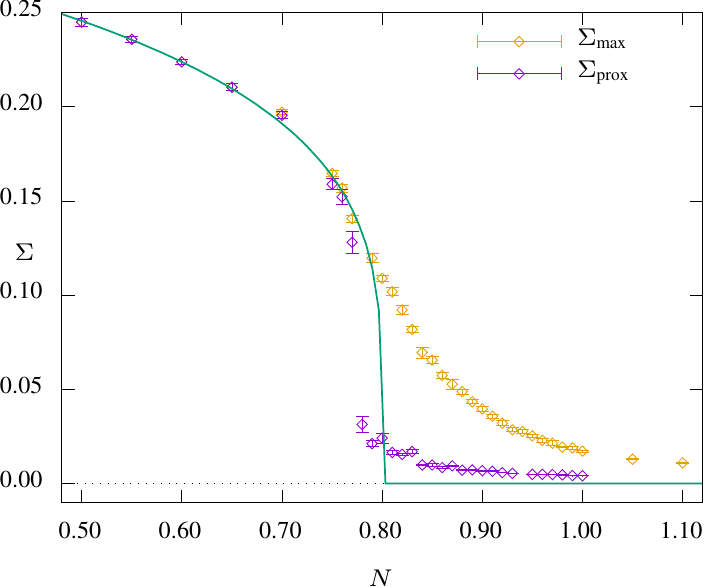}}
	\caption{Chiral condensate in the proximity of the lattice artifact transition $\Sigma_{\text{prox}}$ compared with the maxima $\Sigmam$ and the old fit from Figure~\ref{Fig:Maximum01} (line) for $L=16$.}\label{Fig:cc2}
\end{figure}
To summarize: we conjecture that the critical number
$\Nrc$ separating systems with and without chiral
symmetry breaking and the number $N^\mathrm{t}$
where the two phase transition lines come close or meet 
should be identified.

But how can we explain that
on a finite lattice the chiral condensate is maximal
for $\Nr\geq N^\mathrm{t}$ just to the right of the 
interaction-driven transition line in
Fig. \ref{Fig:peaks} and only vanishes in the infinite
volume limit? 
Without SSB there are two sources for 
a non-zero condensate: the explicit symmetry breaking 
by the fermion mass and fluctuations. 
After (\ref{condensate2}) we have argued that 
\emph{any $v^\mu$-configuration} adds a non-negative
number to the condensate. Near a second order
transition the fluctuations are large and on a finite system these large fluctuations drive the condensate away 
from zero. This explains, why the condensate and the chiral susceptibility\footnote{For $N\geq \Nrc$ there is still a peak of the susceptibility shown as dashed line in Fig.~\ref{Fig:peaks}.} $\chi_\Sigma$ peak near
the interaction-driven transition line.
On the other hand, near a first order transition 
to the lattice artifact phase the fluctuations do not 
necessarily grow and we do not expect
a fluctuation driven condensate. This is the reason,
why for $\Nr\geq N^\mathrm{t}$ (in which case the first-order
and second-order lines are well-separated) 
the condensate is small near the artifact 
transition line and does not depend much on the volume. 
Then we would expect, that the condensate just to
the right of the artifact line is a better approximation
to the condensate in the thermodynamic limit.
In Fig. \ref{Fig:cc2} we plotted for every $\Nr$ the 
maximum of the chiral condensate and its value
in the proximity -- actually just to the right --
of the lattice artifact transition 
line\footnote{Just to the right means three ticks
(in the fixed $\lambda$-grid) to the right.
For comparison we applied this rule to the
maxima of the condensate and the artificial
phase transition line. The points 
where $\Sigma_\mathrm{prox}$
in Fig.~\ref{Fig:cc2} 
have been measured are depicted in Fig.~\ref{Fig:peaks}.}.
We see that for $\Nr<N^\mathrm{t}$ the chiral condensate 
$\Sigma_\mathrm{prox}$ follows the \emph{old} 
fit in Fig.~\ref{Fig:Maximum01} (with the 
form \eqref{eq:critfit} and the parameters from Table~\ref{tab:critfit}). This is expected since
$N^\mathrm{t}\approx\Nrc$. Nevertheless, it further
substantiates our claim that the condensate 
$\Sigma_\mathrm{prox}$ is a better approximation to
the chiral condensate at infinite volume as compared to 
the maximal condensate since fluctuations, which
drive the condensate away from the infinite volume
result, are suppressed.

\section{Conclusions}\label{chap:conclusion}
In the present work we have re-analyzed the long-standing
problem about the critical flavor number in the three-dimensional
(reducible) Thirring models. We used chiral SLAC fermions to have 
full control over the chiral properties of the model. In this 
formulation the chiral and parity symmetry are manifest and 
no fine tuning is required. We reformulated the model 
such that the number of reducible flavors $\Nr$ becomes
a continuous parameter -- offering the possibility of
determining precisely when spontaneous symmetry breaking
ceases to exist. We calculated the maximum of the chiral
condensate, the spectral density and the spectrum of
scalar and pseudo-scalar particles as function of the
flavor number $\Nr$ between $0.5$ and $1.0$.
As a result we find a critical flavor number
\begin{align}
	\Nrc=0.80(4)\,.\label{res:nc}
\end{align}
In particular, we spotted two Goldstone bosons only for
$\Nr\leq \Nrc$.
Since a non-integer value of $\Nr$ probably does not
describe a local quantum field theory (and in particular no
Thirring model),  we conclude that there is no spontaneous 
symmetry breaking in all reducible Thirring models.

With an elaborate and expensive scan of the susceptibility 
related to the interaction term $(\bar{\psi}\Gamma_{\mu}\psi)^2$ as function of the
coupling $\lambda$ and the number of flavors $\Nr$
we spotted -- besides the (probably first order)
ubiquitous lattice artifact transition -- a (probably 
second order) transition for all $\Nr$ greater than
\begin{equation}
N^\mathrm{t}=0.78(4)\,.
\end{equation} 
We gave several arguments why $\Nrc$ and $N^\mathrm{t}$
should be identified. 
Thus
we expect that for an arbitrary number
of flavors there exists a continuous phase transition:
for every $\Nr\leq \Nrc=N^\mathrm{t}$
there is a transition with spontaneous breaking of chiral
symmetry and for every $\Nr\geq \Nrc$ there 
exists a transition \emph{without} spontaneous breaking
of chiral symmetry. But since $\Nrc<1$ only 
the latter transition can be used to construct
continuum Thirring models with $\Nr=1,2,3,\dots$.

The result (\ref{res:nc}) improves the 
result in \cite{Wellegehausen:2017goy} 
and results of lattice Monte-Carlo simulations with
domain wall fermions in \cite{Hands:2018vrd}.
The latter support our 
claim that there is no spontaneous symmetry breaking 
for $\Nr>1$. Since there 
are still major technical issues to be studied in the domain
wall formulation -- such as 
the discrepancy between the bulk and surface formulation and 
the additional $L_s\to \infty$ extrapolation -- the conclusion
for the $\Nr=1$ case is only preliminary. But it seems to
disagree with the results in the present
analysis with SLAC fermions and in
\cite{Wellegehausen:2017goy}.

In parallel to the present work L. Dabelow, H. Gies and B. Knorr
investigated reducible Gross-Neveu-Thirring models in three
dimensions with FRG methods by admitting momentum 
dependent vertices in the flow equation for the scale
dependent effective action \cite{Dabelow:2019sty}.
Their new estimate for $\Nrc$ (obtained with their most strict
criterion) is compatible with ours.

We would like to stress that 
our results are not in contradiction with
those in \cite{Schmidt2016,Wellegehausen:2017goy},
where a breaking of parity
symmetry in models with an odd number of \emph{irreducible}
flavors has been reported. 
The irreducible models are very different from 
the  parity invariant reducible models studied in the present
work and in other more recent publications on 
the three-dimensional Thirring model.

Besides the question about the precise value
of $\Nrc$ we witness a convergence of recent
results obtained with sophisticated functional methods 
and lattice simulations based on chiral fermions. 
So the question arises why earlier attempts with
staggered fermions failed to predict an
acceptable value for $\Nrc$?
It has already been pointed out in
\cite{Wellegehausen:2017goy,Hands:2018vrd}, 
and we would like to stress it once more, that the failure of
staggered fermions to find the correct symmetry
(or even universality class) and phase structure
of $3$-dimensional four-Fermi theories
away from weak coupling, is probably also responsible
for the mismatch between DMF and staggered fermion
results near a conformal fixed point in $3+1$ dimensional
non-Abelian gauge theory \cite{Hasenfratz:2017qyr}.
For strongly coupled (fermion) systems
we should be careful to implement all global
internal symmetries in any discretization.

Simulations of fermion systems are rather time
consuming and an elaborate finite size analysis could
not be accomplished in the present work. For example,
to really decide about the order of the 
interaction-driven phase transition above $\Nrc$
requires further extensive studies. Even more demanding
would it be to extract critical exponents of interest
to decide about the universality class of the 
system at criticality. This would allow for a comparison
with recent results obtained with functional methods.
We hope to report on further progress in these directions 
in the near future.

\section*{Acknowledgments}
\noindent
We are grateful to Shailesh Chandrasekharan, 
Lukas Janssen, Simon Hands, Rajamani Narayanan 
and in particular Holger Gies for 
helpful discussions and comments.
We thank Holger Gies for communicating
to us their FRG-results 
prior to publication. Daniel Schmidt 
contributed considerably 
at an earlier stage of this project.
Our simulations 
were performed on the HPC cluster at the University of Jena.

\appendix
\section{Parity}
\noindent
We choose the parity operation $x\to\tilde x=(x_1,x_2,L-x_3)$,
where $L$ is the extend of the box in $3$-direction.
The auxiliary vector field transforms as
\begin{equation}
\tilde v_{1,2}(x)=v_{1,2}(\tilde x),\;\;
\tilde v_3(x)= -v_3(\tilde x)
\end{equation}
and a $2$-component spinor field $\chi$ as
\begin{equation}
\tilde{\chi}(x)=\gamma_3\chi(\tilde x)\,.\label{paritymodes}
\end{equation}
Now it follows at once that if $\chi$ is an eigenfunction of
$\ii\slashed{D}_v$ with eigenvalue $\lambda$, then
$\tilde{\chi}$ is an eigenfunction of $\ii\slashed{D}_{\tilde v}$
with eigenvalue $-\lambda$. 
\subsection{Spectral density}
The invariance under parity implies particular properties
of the spectral density defined in (\ref{spectraldensity})
and the fermion Greenfunction $\Delta$ of the 
reducible Diracoperator in (\ref{diracop2}).
For example, we conclude
\begin{equation}
\rho_{v}(E)=\rho_{\tilde v}(-E)\,.
\end{equation}
Since the space-time integrals over $v_\mu^2$ and $\tilde v_\mu^2$ 
are equal and in addition only the square of 
$\slashed{D}$ enters the effective action, we see
that the latter is parity invariant,
\begin{equation}
S_\mathrm{eff}(v)=S_\mathrm{eff}(\tilde v)\,.
\end{equation}
Since the averaging over the auxiliary field is done
with the parity invariant factor $e^{-S_\mathrm{eff}}$
we end up with the relation (\ref{av_density})
which states, that the averaged spectral density 
$\bar\rho(E)$ is an even function of
the spectral parameter.
\subsection{Fermion Greenfunction}
The fermion Greenfunction of the reducible system is
\begin{equation}
\Delta=\frac{1}{D}
=\begin{pmatrix}
\Delta_+&0\\ 0&\Delta_{-}
\end{pmatrix},\quad 
\Delta_\pm=\frac{1}{\ii\slashed{D}\pm\ii m}\,,
\end{equation}
where $\slashed{D}$ belongs to the
irreducible system.
The Greenfunction is a linear combination 
of $\sigma_0$ and $\sigma_3$, see 
(\ref{decomp_green1}) and (\ref{decomp_green2}).
It follows that
\begin{align*}
&
\frac{1}{4}
\trf\big(\sigma_a\Delta_{xx}\big)
\trf\big(\sigma_b\Delta_{yy}\big)\\
&=
\begin{pmatrix}
A_{xx}A_{yy}&0&0
&A_{xx} B_{yy}\\
0&0&0&0\\
0&0&0&0\\
B_{xx}A_{yy}&0&0
&B_{xx}B_{yy}
\end{pmatrix}\,,
\end{align*}
where $\trf$ denotes the trace in flavor space,
and in addition
\begin{align*}
&\frac{1}{2}\trf\big(\sigma_a\Delta_{xy}\,\sigma_b\Delta_{yx}\big)\\
&=A_{xy}A_{yx}\begin{pmatrix}1&0&0&0\\ 0&1&0&0\\
0&0&1&0\\0&0&0&1
\end{pmatrix}
+B_{xy}B_{yx}\begin{pmatrix}1&0&0&0\\ 0&-1&0&0\\
0&0&-1&0\\0&0&0&1
\end{pmatrix}
\\
&+A_{xy}B_{yx}\begin{pmatrix}0&0&0&1\\ 0&0&\ii&0\\
0&-\ii&0&0\\1&0&0&0
\end{pmatrix}
+B_{xy}A_{yx}\begin{pmatrix}0&0&0&1\\ 0&0&-\ii&0\\
0&\ii&0&0\\1&0&0&0
\end{pmatrix}\,.
\end{align*}
Recall, that the correlation matrix (\ref{corrmatrix})
involved suitable traces over the spinor indices
as well, $\tr=\trd\trf$.
Next we study the transformation of the Greenfunction
under parity. Since the eigenmodes change 
according to (\ref{paritymodes}) and the eigenvalues
swap signs, we have
\begin{equation}
\Delta_\pm(x,y,v)=-\gamma_3 \Delta_\mp(\tilde x,\tilde y,
\tilde v)\gamma_3\,,\label{paritygf1}
\end{equation}
which in turn implies
\begin{equation}
A_{xy}(v)=-\gamma_3 A_{\tilde x\tilde y}(\tilde v)\gamma_3,\quad 
B_{xy}(v)=\gamma_3 B_{\tilde x\tilde y}(\tilde v)\gamma_3\,.
\end{equation}
It follows, for example, that
\begin{equation}
\sum_\vx \trd A_{xx}(v)
=-\sum_\vx \trd A_{\tilde x\tilde x}(\tilde v)
=-\sum_{\vx} \trd  A_{x x}(\tilde v)\,.
\end{equation}
In the last step we used, that the two $\gamma_3$ in the
conjugation (\ref{paritygf1}) chancel 
under the trace over Dirac indices and that
summing over all $\vx$ is the same as 
summing over all $\tilde{\vx}$.
Averaging with the parity-invariant effective action
over the auxiliary field results into
\begin{equation}
\sum_{\vx} \trd\,\langle A_{xx}\rangle=
-\sum_{\vx} \trd\,\langle A_{xx}\rangle=
0\,.
\end{equation}
Similarly one obtains
\begin{equation}
\sum_{\vx,\vy} \big\langle \trd A_{xx}B_{yy}
\big\rangle=0=
\sum_{\vx,\vy} \big\langle \trd A_{xy}B_{yx}
\big\rangle\,.
\end{equation}
It follows that the correlation matrix $C(t)$ is diagonal with eigenvalues $C_a(t)$ given in (\ref{corr_matrix1}).
Finally note that $\,\ii\trd\langle B_{xx}\rangle$ is just the
chiral condensate $\Sigma$.

\section{Numerical Differentiation}
\noindent
While numerical differentiation of smooth data is easily done 
by discrete derivative stencils, non-smooth and particularly
noisy data is hard to differentiate numerically. This is seen 
in Figure~\ref{Fig:dk} where the markers show the result of
applying the stencil
\begin{align}
\partial_\lambda k (\lambda_{i}) = \frac{k(\lambda_{i+1})-k(\lambda_{i-1})}{2\delta\lambda} + \mathcal{O}(\delta\lambda^2)
\end{align}
to the rather smooth looking data of Figure~\ref{Fig:k}. Particularly, in the interesting regime around $\Nr=0.8$ such a numerical differentiation is basically useless because of the large noise. Another approach, which we will use in the following, is total-variation (TV) regularized differentiation \cite{chartrand2011}. It reformulates the problem as a global optimization problem such that the minimum of the functional
\begin{align}
F(u) = \| I(u) - (k - k(\lambda_0)) \| + \alpha R(u)
\end{align}
is assumed for an approximation $u\approx \partial_\lambda k$. 
Here $I(u)$ is an (appropriate discrete) integration
operation and $\|\cdot\|$ an appropriate norm such that $k$ is
obtained from integrating its derivative $u$. Afterwards a
regulator term $R$ can be added to smooth the minimizing solution
$u$.\footnote{Details of the implementation can be found in
\cite{chartrand2011}. We used the python translation from
\url{https://github.com/stur86/tvregdiff} of their MATLAB code
with minor modifications.} While one can clearly see the
smoothing behavior of this approach in Figure~\ref{Fig:dk}, the
important information about the peak (location) is not distorted
compared to the naive scheme. We always cross-checked that the TV 
result was plausible within the naive scheme; however, we cannot
assign a pointwise uncertainty to the TV result due to the global
procedure for obtaining it.

\section{Monte-Carlo simulations}
\noindent
For performance reasons, the Monte-Carlo simulations have been performed in the two-component irreducible representation. In the HMC algorithm we compute the fermion determinant by introducing $p$ pseudo-fermions $\phi,\phi^\dagger$
\begin{equation}
\begin{aligned}
&\det \left(D^\dagger D\right)^\frac{\Nr}{2}=\det \left(D_+^\dagger D_+\right)^\frac{\Ni\,p}{2p}\\
\sim &\int \mathcal{D}\phi \mathcal{D}\phi^\dagger\,\exp\left\lbrace-\sum \limits_p\phi_p^\dagger r\left(D_+^\dagger D_+,\frac{\Ni}{2}\right)\phi_p\right\rbrace\,,
\end{aligned}
\end{equation}
where the function $r(A,n)$ is a rational approximation of the inverse fermion matrix
\begin{equation}
\left(D_+^\dagger D_+\right)^{-{\frac{\Ni}{2p}}}\approx r\left(D_+^\dagger D_+,\frac{\Ni}{2p}\right)
\end{equation}
and
\begin{equation}
r(M,k)=a_0(k)+\sum \limits_{i=1}^n \frac{\alpha_i(k)}{M+\beta_i(k)}.
\end{equation}
The coefficients $\alpha_i$ and $\beta_i$ depend on the degree $n$ of
the approximation, on the power of the inverse fermion matrix $k$
and on the spectral range of $M$. Details on the RHMC algorithm can
be found in \cite{Clark:2003na}. In this way we are able to perform lattice
simulations for any rational flavor number. To speed up the
simulations, we use different approximations in the HMC
trajectory and the metropolis acceptance step. For most of our
simulations we use $p=4$ pseudo-fermions and a degree of the
approximation $n_\text{HMC},n_\text{acc}=10,25$.

The inverse of the shifted fermion matrix in the rational approximation is computed by a multi-mass conjugate gradient (CG) solver. During the CG iterations, we have to apply the SLAC operator to a pseudo-fermion field. Here, we make use of a special property of the SLAC derivative: It is diagonal in momentum space and we obtain
\begin{equation}
\begin{aligned}
(D_+ \phi)(x)=&\iFT\left[\sum \limits_p \ii \slashed{p} \, FT[\phi](p)\right](x)\\
&+\left(\ii \gamma_\mu v_\mu(x)+m\right)\phi(x)
\end{aligned}
\end{equation}
where the sum is over all lattice momenta $p$. Instead of using a three-dimensional (parallelized) Fourier transformation, we apply one-dimensional Fourier transformations that are computed in parallel. Although there is communication overhead, this method is on small lattices far more efficient than a three-dimensional Fourier transformation.

\renewcommand{\eprint}[1]{ \href{http://arxiv.org/abs/#1}{[arXiv:#1]}}

\bibliography{main}

\end{document}